\documentclass[bibyear]{aa}  
\pdfoutput=1
\usepackage{graphicx,longtable}
\usepackage{natbib}
\usepackage{txfonts}
\usepackage[normalem]{ulem}

\begin{document}

   \title{High-precision polarimetry of nearby stars ($d<50$ pc)}

   \subtitle{Mapping the interstellar dust and magnetic field inside the Local Bubble }

   \author{V. Piirola
          \inst{1,2}
          A. Berdyugin
          \inst{1} 
          P. C. Frisch
          \inst{3}
          M. Kagitani
          \inst{4}
          T. Sakanoi
          \inst{4}
\\
         S. Berdyugina
          \inst{5}
         A.A. Cole
         \inst{6}
         C. Harlingten
         \inst{6}
          \and 
         K. Hill
         \inst{6}           
          }

   \institute{Department of Physics and Astronomy,  University of Turku, Finland 
 \\
              \email{piirola@utu.fi, andber@utu.fi}
         \and
              FINCA, University of Turku, FI-20014 Turku, Finland
       \and
             Department of Astronomy and Astrophysics, University of Chicago, Chicago IL 60637 USA
       \and
             Graduate School of Science, Tohoku University,
              Aoba-ku, Sendai 980-8578 Japan 
      \and
             Leibniz-Institut f\"{u}r Sonnenphysik, D-79104 Freiburg, Germany
       \and
       Greenhill observatory, School of Natural Sciences, University of Tasmania, Private Bag 37, Hobart, TAS 7001 Australia           
             }

   \titlerunning{High-precision linear polarimetry of nearby stars}
   \authorrunning{V. Piirola et al. }

   \date{Received 2019 December 16 ; accepted 2020 January 27 }

 
  \abstract
   {We investigate the linear polarization produced by interstellar dust aligned by the magnetic 
   field in the solar neighborhood ($d< 50$ pc). We also look for intrinsic effects 
  from circumstellar processes, specifically in terms of polarization variability and 
wavelength dependence.} 
   { We aim to detect and map dust clouds which give rise to statistically significant amounts of
polarization of the starlight passing through the cloud, and to determine the interstellar magnetic 
field direction from the position angle of the observed polarization. }
   { High-precision broad-band ($BVR$) polarization observations are made of 361 stars
in spectral classes F to G, with detection sensitivity at the level  of or better than $10^{-5}$ (0.001\%). 
The sample consists of 125 stars in the magnitude range 6-9 observed at the 2.2 m UH88
telescope on Mauna Kea, 205 stars in the magnitude range 3-6 observed at the Japanese (Tohoku) 
T60 telescope on Haleakala, and 31 stars in the magnitude range 4-7 observed at the 1.27 m
 H127 telescope of the Greenhill Observatory, Tasmania. Identical copies of the Dipol-2 polarimeter
 are  used on these three sites. 
}
   { Statistically significant ($>3 \sigma$) polarization is found in 115 stars, and 
$> 2 \sigma$ detection in 178 stars, out of the total sample of 361 stars. Polarization maps
based on these data show filament-like patterns of polarization position angles, which are related to both the
heliosphere geometry, the kinematics of nearby clouds, and the Interstellar Boundary EXplorer (IBEX)  
ribbon magnetic field. From long-term multiple observations, a number ($\sim$ 20) of stars show 
evidence of intrinsic variability at the $10^{-5}$ level. 
This can be attributed to circumstellar effects (e.g., debris disks and chromospheric activity). The star HD 101805  
shows a peculiar wavelength dependence, indicating size distribution of scattering particles different from that 
of a typical interstellar medium.  Our high-S/N  measurements of  nearby stars with very low polarization also 
provide a useful dataset for calibration purposes. }
   {}


   \maketitle
%

\section{Introduction}
 
In the course of polarization studies of astrophysical objects that require very high
S/N \citep[see][]{Berdyugin16, Berdyugin18}, we observed samples of nearby stars ($d < 50$ pc)
in order to determine and subtract the instrumental polarization produced by the telescope. This is achieved 
with a typical uncertainty of 2-3 x $10^{-6}$ (ppm) for each run, normally spanning intervals of between 
several days and one month. To avoid intrinsic effects as much as possible, stars of spectral types F-G were 
selected. These calibration measurements are crucial for each of the research programs, but they also
provide a valuable database of the minute amounts of polarization in each of the observed 
stars. The individual stellar data points can in turn be used to map the interstellar magnetic field
and dust content along the path that the observed starlight has traversed.

While the interstellar magnetic field (ISMF) structure and dust content are relatively well understood
from the polarization maps based on measurements of stars at larger distances -- $d >$ 50 pc up to 
the kiloparsec ranges \citep[see e.g.,][]{Berdyugin14} -- the very low dust content and therefore 
low degree of polarization within the Local Bubble have prevented detailed polarization studies
until the recent development of extremely high-S/N polarimeters, with  $10^{-6} - 10^{-5}$  detection
sensitivity \citep[see e.g.,][]{Bailey10, Bailey15, Piirola14, Cotton17, Cotton19}.

Encouraged by the successful implementation of our high-precision polarimeter (Dipol-2),
and the availability of sufficient amounts of observing time from remotely operated telescopes
at good observing sites, we have initiated a dedicated program for deriving the structure
of the very local magnetic field from stellar polarization data. The need for a survey of optical 
polarizations that trace the local interstellar magnetic field became clear with the discovery 
by Interstellar Boundary EXplorer (IBEX) of an arc (or "ribbon") of energetic neutral atoms 
whose center traces the direction of the ISMF shaping the 
heliosphere \citep{McComas09, Schwadron09}.

Target stars for this program were selected from objects in the 
Hipparcos catalog (Perryman et al. 1997).  Channels devoid of nearby suitable target
objects were found in the Hipparcos data in many locations, indicating either
an irregular distribution of dust or an irregular distribution of suitable
target stars. Some earlier results from our nearby star observations have been included in the studies 
by \citet{Frisch2015, Frisch15}, tracing the structure of the very local ISMF. 

In the present paper we give a detailed description of our current dataset and
discuss the results based on the polarization map obtained. The statistical significance of the
detections is addressed. It is also interesting to look for evidence of intrinsic 
polarization variability in some of the stars observed. Examples are given of particular
wavelength dependence found, suggesting effects from a circumstellar debris disk.
 
\section{Observations}

We carried out observations in 2014-2019 at three telescopes, the 2.2 m UH88 telescope
on Mauna Kea, the Tohoku 60 cm telescope (T60) on Haleakala, and the University of Tasmania (UTAS)
1.27 m (H127) telescope at  Greenhill Observatory, Tasmania.
Observations were made with the simultaneous three-color ($BVR$) polarimeter
Dipol-2  \citep{Piirola14}. Identical copies of the instrument are used at each of
the three sites. At UH88 and T60 the observations were carried out in remote
operation mode. Some additional data on the stars in our sample were
obtained at the Nordic Optical Telescope (NOT) and William Herschel (WHT) telescopes at ORM, La Palma,
A summary of the observations is given in Table 1.

The polarimeter, Dipol-2, is capable of making simultaneous
measurements in three passbands, $B, V$, and $R$ (see Figs. 1-2, and Table 2), with high 
sensitivity. The detection limit of polarization is at the level of $10^{-5}$, which is set in practice
by photon noise. An important asset of the instrument is that the sky
background polarization is directly (optically) eliminated. The perpendicularly
polarized components of sky are superimposed by the
plane parallel calcite beam splitter, and sky polarization is thereby
canceled \citep{Piirola73}. This is essential, as the polarized flux from
scattered skylight can exceed the signal
from the target by orders of magnitude, particularly in bright Moon conditions. Dipol-2
has been found to be a very stable and reliable instrument as demonstrated
recently by detection of the variable polarization at the 0.1\%\ level from the massive binaries 
\object{HD 48099} (Berdyugin et al. 2016) and \object{$\lambda$ Tauri} (Berdyugin et al. 2018).
%
   \begin{figure}
   \centering
   \includegraphics[width=\hsize]{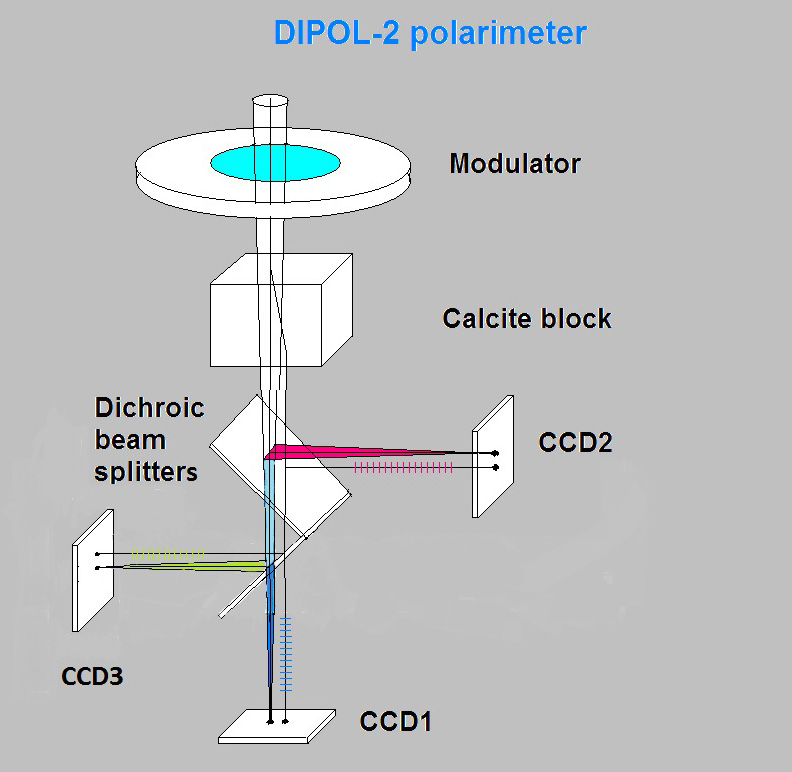}
      \caption{Scheme of the Dipol-2 polarimeter. Rotatable superachromatic $\lambda$/2 retarder plate modulates the relative intensities of the two polarized beams produced by the calcite crystal, with an amount proportional to the degree of linear polarization of the incoming radiation. Two dichroic mirrors split the light into three passbands: blue, visible, and red. The fluxes of the two polarized stellar images in each band are measured with three highly sensitive cooled CCD detectors. 
              }
         \label{Fig1}
   \end{figure}
%
   \begin{figure}
   \centering
   \includegraphics[width=\hsize]{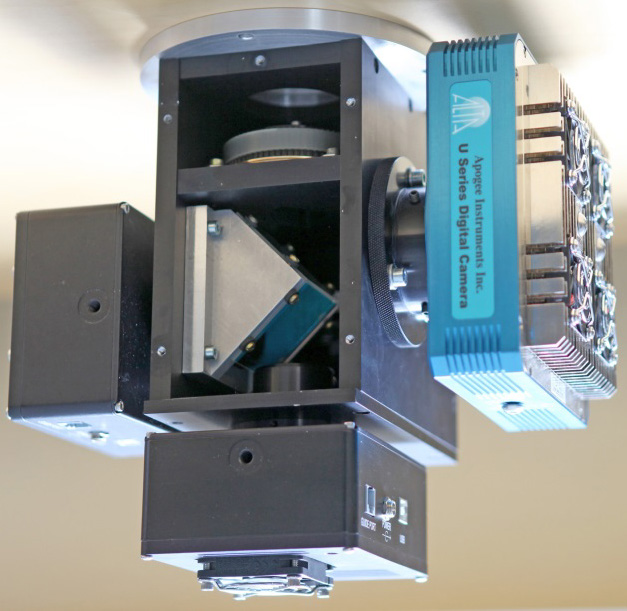}
      \caption{ Dipol-2 polarimeter with the front cover removed, showing (from top) the polarization modulator unit (retarder plate and the calcite plate), the dichroics unit dividing the light onto three CCDs: the blue (right), the visible (bottom), and the red (left).   
              }
         \label{Fig2}
   \end{figure}
%
%

The Dipol-2 polarimetry routine consists of cycles of 16 exposures
at different orientations of the superachromatic half-wave
retarder (22.5$\degr$ steps), corresponding to a full (360$\degr$) rotation of the
retarder. Each group of four successive exposures gives one independent measurement
of the normalized Stokes parameters $q=p\cos 2\theta$ and $u=p\sin 2\theta $,
where $p$ is the degree of linear polarization and $\theta$ the position angle of
the maximum electric vector, in the equatorial frame of reference. 
Accordingly, one cycle gives four independent measurements of $q$ and
$u$.  

For the highest S/N measurements it is advantageous to strongly defocus images
to spread light over a very large number of pixels. In this way we can expose up to
$10^8$ electrons in one stellar image without saturating the CCD pixels. Even if there
are minor shifts in the position of the two perpendicularly polarized images of the
target, the vast majority of the pixels remain the same over the full measurement cycle
(16 exposures). Minor imperfections in the flat field are eliminated in the reductions,
because the ratio of the o- and e- beam transmission, and efficiency, if constant, is
automatically canceled in the reduction algorithm. This provides inherently very
stable instrument and detection sensitivity better than $10^{-5} (< 10$ ppm) in $\sim$ 1 hour
for sufficiently bright stars. Moreover, the Dipol-2 polarimeter is photon noise limited
down to these very low polarization signal levels.

In order to reduce the photon noise to the required ($10^{-5}$) level,
the total telescope time used for the nightly observation of each star was 0.5-1.5 hours,
depending on the brightness of the star. With a typical single exposure time of 1-3 s,
the total number of individual observations of the normalized Stokes parameters $q$ 
and $u$ for one star was usually in the range 128-256. This provides very good 
statistical error estimates for the nightly average points of $q$ and $u$ for each star. 

Standard CCD reduction procedures (bias and dark subtraction,
flat fielding) were applied prior to extracting the fluxes from
the double images of the target formed on the CCD by the polarizing
calcite beam splitter. A special centering algorithm and subframing
procedures were used to facilitate the processing of a large number --
up to several hundred -- of exposures at the same time. In computing
the mean values of $q$ and $u$ we applied a "$2\sigma$" iterative
weighting algorithm. The initial mean and standard deviation were
obtained by applying equal weights to all points. Subsequently, on each step,
individual points deviating by more than two standard deviations from
the mean ($d > 2\sigma$) were given a lower weight proportional to the
inverse square of the error estimate, $e_x$. The value $e_x = \sigma$  for $d < 2\sigma$ 
was assumed to increase linearly from $e_x = 1\sigma$ to 3$\sigma$ with $d$ increasing
from $2\sigma$ to $3\sigma$. Points with $d > 3\sigma$ were rejected. The
procedure converges fast and values of mean and standard deviation
are obtained within a few iterations. Under normal conditions, 6 --  8
\%\ of individual points deviated by more than $2\sigma$  and were given
lower weight ($W <1$). The remaining 92 -- 94 \%\ of points were
equally weighted ($W = 1$). The weighting procedure helps to suppress
effects from transient clouds, moments of bad seeing, cosmic
ray events, and so on.

%
\begin{table}
\caption{Summary of observations.}             
\label{table1}      
\centering                          
\begin{tabular}{c c c c}        
\hline\hline                 
            \noalign{\smallskip}
Telescope &  JD Interval  &Stars & mag range \\    
\hline                        
            \noalign{\smallskip}
  UH88 & 2456818-7678 & 125 & 6.1 -- 9.1 \\      
   T60 & 2456994-8508 & 205 & 3.9 -- 6.5 \\
  H127 & 2457775-8183 & 31 & 4.1 -- 6.9 \\
          & Additional data: & & \\
  WHT & 2457206-7407 & 12 & 7.2 -- 8.4 \\
  NOT & 2458687-8688 & 15 & 3.8 -- 5.8 \\
\hline                                   
\end{tabular}
\end{table}
%
%
\begin{table}
\caption{Equivalent wavelengths and full widths at half maximum (FWHM) of the Dipol-2 passbands}     
\label{table:2}      
\centering                          
\begin{tabular}{c c c}        
\hline\hline                 
            \noalign{\smallskip}
Passband& $\lambda_{eq}$(nm) & FWHM (nm)  \\    
\hline                        
            \noalign{\smallskip}
 $B$ & 450 & 110  \\      
 $V$ & 545 & 89 \\
 $R$ & 655 & 120 \\
\hline                                   
\end{tabular}
\end{table}
%

The simultaneous polarization measurements in the three ($BVR$) passbands provided by
Dipol-2 are very useful for studying the wavelength dependence of polarization. Indeed
they also improve the efficiency. There is only a small amount of internal absorption in the dichroic beam
splitters used to separate the color passbands.  

In the case of the extremely
low polarization values found in the stars inside the Local Bubble, the S/N in each of
the wavelength bands ($BVR$) may not be sufficient to obtain useful data for constraining the 
wavelength dependence of interstellar polarization, and thereby the aligned grain size distribution.
In such cases, and because of the relatively flat shape of the IS polarization curve in
the optical part of the spectrum, it is meaningful to compute `broad-band' (400-800 nm)
polarization values by weighted averaging of the normalized Stokes parameters $q$ and $u$
obtained in the $B$, $V$, and $R$ passbands to improve the S/N and the statistical significance of
the detection for very weak polarization signals.

   \begin{figure*}
   \centering
    \includegraphics[width=16.5cm]{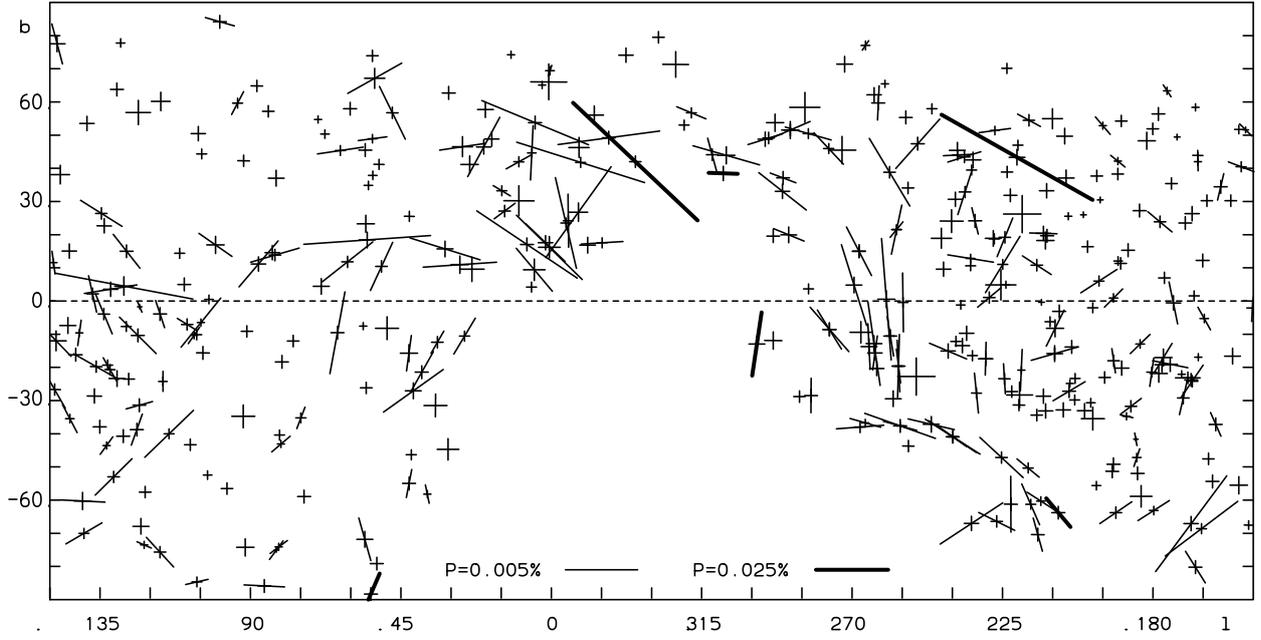}
   \caption{Polarization map based on the observed sample of nearby stars ($d<50$ pc),
      plotted in Galactic coordinates. The length of the bars is proportional to the degree of
      polarization and the orientation gives the direction of the maximum electric vector.
      Regions with aligned polarization vectors can be seen, suggesting filament-type structures. 
      Two different polarization scales are used for clarity, as indicated in the bottom of the panel.   
      For  low observed degrees of polarization, $p<2\sigma$ (no detection), only a cross with 
      bar lengths, $\sigma$, is plotted.}

              \label{Fig3}%
    \end{figure*}
%
%
   \begin{figure*}
   \centering
    \includegraphics[width=16.5cm]{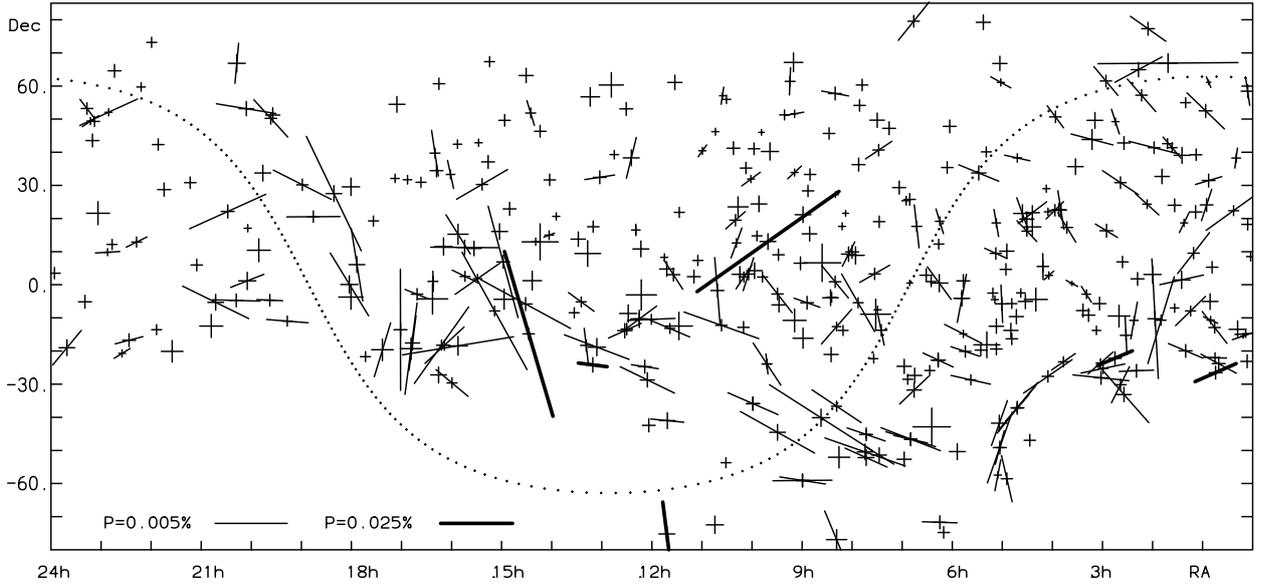}
   \caption{Polarization map based on the sample of nearby stars ($d<50$ pc), plotted in Equatorial coordinates.   
      Two different polarization scales are used for clarity, as indicated in the bottom of the panel. For  
      low observed degrees of polarization, $p<2\sigma$ (no detection), only a cross with 
      bar lengths, $\sigma$, is plotted. The dotted line shows the Galactic equator.}

              \label{Fig4}%
    \end{figure*}

\section{Results and Discussion}

\subsection{Instrumental polarization}

Because of the very small degree of polarization produced by the interstellar dust in the Solar vicinity,
it is crucial to determine and subtract the effects of the telescope optics and the instrument itself from
the measured polarization. This is done by observing nearby stars which can be assumed to be almost
 unpolarized thanks to their proximity to us and freedom from intrinsic polarization effects ensured by the selection of
the spectral type (F-G main sequence stars). 

However, we cannot assume that any single star has zero
polarization within the strict limits required for the present work. Therefore, we always observe a
sample of at least 15-20 stars in different parts of the sky and determine the instrumental polarization
as the average of the $q$ and $u$ from the sample. In this way, weak effects of interstellar
polarization in each of the observed stars tend to cancel out, because the ISMF
direction varies strongly in different parts of the sky. Possible intrinsic (circumstellar or photospheric)
effects are also likely to be randomly oriented within the sample.

In Tables 3-5 we list the values of the instrumental polarization, $q_t$ and $u_t$, obtained
for the UH88, T60, and H127 telescopes, respectively, for each run carried out. Statistical uncertainties
are typically in the range $2-3 \times 10^{-6}$, and the values for each telescope do not change significantly from
one run to another. This is reassuring, since samples in different parts of the sky are used for
different runs (seasons), and possible residual effects from interstellar polarization appear to be
small. Obviously, there are slow long-term drifts and small jumps from mirror cleaning or other
effects from the telescope optics.

We excluded
stars with significant  detected polarizations  from the instrumental polarization computation, and applied a
simple iterative procedure to calculate the average $q$ and $u$ from the sample to provide $q_t$ 
and $u_t$ in each of the $BVR$ passbands separately. These values, $q_t$ and $u_t$, are then 
subtracted from the observed $q$ and $u$ in each passband for all of the observed stars. The telescopes 
we use for the present work (UH88, T60, and H127) are equatorially mounted, which is convenient in the 
sense that the telescope polarization does not rotate on the sky. It only gives a constant shift in $(q,u)$.

The superachromatic half-wave plates we use have very good efficiency throughout the whole
wavelength range of the Dipol-2 passbands (400-800 nm). Nevertheless, we observed high
polarization standard stars to check for the polarization scale calibration, and for the zero-point of 
position angles.

In each run, at least two different large polarization standards were observed, 
typically both at the beginning and at the end of the run. The stars \object{HD 25443},
\object{HD 161056},  \object{HD 204827}, \object{BD+25 727}, and \object{BD+59 389} were used for this purpose.
We found evidence of small calibration coefficients (1.02 -- 1.04) needed in the $V$ and $R$ passbands. 
Though these differences can be partially due to systematic errors in the published values 
\citep{Hsu82, Turnshek90, Schmidt92}, we applied the corrections 
to bring our data into the system commonly used by other investigators. In any case, for the very
small polarization degrees found in the present study, these scale corrections are
entirely negligible.

%
\begin{table}
\caption{Instrumental polarization of the UH88 telescope in the $B$, $V$, and $R$ passbands,  in units of 
$10^{-6}$ (ppm), for  the observing runs in 2014-2016. The standard errors are in the range 2-4 ppm for each run.}             
\label{table2}      
\centering                          
\begin{tabular}{c r r r r r r r r}        
\hline\hline                 
            \noalign{\smallskip}
 JD Interval & $q_{tB}$ & $u_{tB}$ & & $q_{tV}$ & $u_{tV}$ & & $ q_{tR}$ & $u_{tR}$\\    
\hline                        
            \noalign{\smallskip}
2456818-6825 & 180&-231 & & 57& -112 & & 24& -74\\     
2456849-6857 & 198&-258 & & 43 & -118 & & 16& -72\\
2457284-7291 & 189&-247 & & 48& -131 & & 35& -86\\
2457413-7420 & 187&-252 & & 64& -121 & & 41& -82\\
2457431-7446 & 199&-227 & & 57& -111 & & 37& -76\\
2457556-7568 & 199&-243 & & 53& -127 & & 36& -84\\
2457645-7657 & 182&-251 & & 55& -138& & 34& -91\\
2457671-7681 & 191&-238 & & 56& -130& & 34& -99\\ 
\hline                                   
\end{tabular}
\end{table}
\begin{table}
\caption{ Instrumental polarization of the T60 telescope in the $B$, $V$, and $R$ passbands,  in units of $10^{-6}$ (ppm), 
for the observing runs in 2014-2018.
 The standard errors are in the range 2-3 ppm for each run.}
\label{table3}      
\centering                          
\begin{tabular}{c r r r r r r r r}        
\hline\hline                 
            \noalign{\smallskip}
JD Interval & $q_{tB}$ & $u_{tB}$ & & $q_{tV}$ & $u_{tV}$ & & $ q_{tR}$ & $u_{tR}$\\    
\hline                        
            \noalign{\smallskip}
2456994-6999 & 34& 11& & 27& -3& & 44& -4\\
2457039-7047 &-7 &-7& &2&-5& &10&-6\\
2457154-7172 &3&-14& &6&-5& &17&-11\\
2457205-7256 &-32&-27& &-30&-2& &-29&2\\
2457297-7304 & -7&-4& &-25& 3& &-22&18\\
2457354-7370 &  1& -15& & -25& -1& & -39& 12\\
2457490-7503 & -2& -22& & -24&-13& & -31& -2\\
2457686-7701 &-24& -26& & -36&  3& & -47& 13\\
2457774-7786 &-11& -23& & -30&  1& & -53&  5\\
2457894-7895 &-11& -30& & -12& -9& & -19&  2\\
2458025-8113 &-27& -20& & -27&  4& & -29&  9\\
2458134-8196 &-12& -19& & -24&  5& & -30& 15\\
2458316-8352 &-20& -26& & -34&  5& & -50&-11\\
2458406-8452 &-24& -13& & -40&  9& & -49& 28\\
2458455-8508 &-16& -26& & -35& -6& & -47&  1\\
\hline                                   
\end{tabular}
\end{table}
%
%
\begin{table}
\caption{Instrumental polarization of the  H127 telescope in the $B$, $V$, and $R$ passbands,  in units of 
$10^{-6}$ (ppm), for  the observing runs in 2017 and 2018. The standard errors are in the range 2-4 ppm for 
the first run, and 4-8 ppm for the second run.
The change in instrumental polarization from 2017 to 2018 is due to operations with the main mirror.}    
\label{table2}      
\centering                          
\begin{tabular}{c r r r r r r r r}        
\hline\hline                 
            \noalign{\smallskip}
 JD Interval & $q_{tB}$ & $u_{tB}$ & & $q_{tV}$ & $u_{tV}$ & & $ q_{tR}$ & $u_{tR}$\\    
\hline                        
            \noalign{\smallskip}
2457775-7799 & -90& -3& & -49& -9 & & -43& -1\\     
2458125-8183 & 175& 207& & 81& 108& & 48& 90\\
\hline                                   
\end{tabular}
\end{table}
%
%

%
%
\setcounter{table}{7}
\begin{table*}
\caption{  Broad-band (400-800 nm) polarimetric data of nearby stars observed
   at the H127 telescope. The normalized Stokes parameters, $q$ and $u$, and the degree of polarization, 
  $p$, are given in units of $10^{-6}$ (ppm), and the position angle $\theta$ in the Equatorial frame of references. }  
\label{table:7}      
\centering          
\begin{tabular} {rcrrrrrlrrrlrl}
\hline\hline
  \noalign{\smallskip}
HD &  R.A. & Decl.    & $l(\degr)$  &  $b(\degr)$ & $V_{mag}$  &  Par. & Sp & $q$ &  $u$ & $p$ &$\pm e_p$ & $\theta(\degr)$ &$\pm e_\theta (\degr)$  \\    
  \noalign{\smallskip}
\hline
            \noalign{\smallskip}
  28454 &  04 27  06.0 & -46 56 51 & 253.18 & -43.71 &  6.10 & 30.54 &F5.5V &    0 &    1 &   1 &$\pm$  8 &  56.7 &$\pm$ 41.2 \\
  29992 &  04 42  03.5 & -37  08 40 & 239.92 & -40.91 &  5.05 & 34.75 &F3VI  &    8 &  -38 &  40 &$\pm$  7 & 140.8 &$\pm$  5.0 \\
  31746 &  04 54 53.0 & -58 32 52 & 267.62 & -37.96 &  6.11 & 32.82 &F5V   &   29 &   14 &  33 &$\pm$  8 &  13.1 &$\pm$  7.1 \\
  32743 &  05  02 48.7 & -49  09  05 & 255.64 & -37.63 &  5.37 & 38.22 &F5V   &   21 &  -12 &  24 &$\pm$  5 & 164.9 &$\pm$  5.3 \\
  32820 &  05  03 54.0 & -41 44 42 & 246.33 & -37.15 &  6.30 & 31.51 &F8V   &   25 &  -21 &  33 &$\pm$ 11 & 160.4 &$\pm$  9.7 \\
  33262 &  05  05 30.7 & -57 28 22 & 266.03 & -36.72 &  4.71 & 86.02 &F9V   &   20 &   -8 &  22 &$\pm$  5 & 169.6 &$\pm$  5.9 \\
  38858 &  05 48 34.9 &  -04  05 41 & 209.38 & -15.84 &  5.97 & 65.55 &G2V   &   31 &  -12 &  34 &$\pm$ 11 & 169.2 &$\pm$  9.0 \\
  40105 &  05 54 10.8 & -50 21 45 & 257.71 & -29.44 &  6.52 & 28.00 &K1IV  &   21 &    5 &  22 &$\pm$ 11 &   7.0 &$\pm$ 13.6 \\
  43834 &  06 10 14.5 & -74 45 11 & 285.76 & -28.80 &  5.09 & 97.90 &G7V   &   -8 &   -9 &  12 &$\pm$  8 & 114.2 &$\pm$ 16.9 \\
  44447 &  06 15 06.0 & -71 42 10 & 282.28 & -28.49 &  6.62 & 30.71 &G0V   &  -24 &    1 &  24 &$\pm$  9 &  88.3 &$\pm$ 10.6 \\
            \noalign{\smallskip}
  45289 &  06 24 24.4 & -42 50 51 & 250.79 & -22.76 &  6.67 & 35.87 &G2V   &   -1 &   28 &  28 &$\pm$ 26 &  46.3 &$\pm$ 21.8 \\
  49095 &  06 45 22.9 & -31 47 37 & 241.29 & -15.08 &  5.92 & 41.65 &F6.5V &   -1 &  -28 &  28 &$\pm$ 10 & 133.6 &$\pm$  9.9 \\
  50223 &  06 49 54.6 & -46 36 52 & 256.12 & -19.60 &  5.14 & 39.70 &F5.5V &  -25 &   16 &  30 &$\pm$  8 &  73.9 &$\pm$  7.4 \\
  52298 &  06 57 45.4 & -52 38 54 & 262.63 & -20.34 &  6.94 & 26.98 &F8V   &    6 &   18 &  19 &$\pm$ 10 &  35.1 &$\pm$ 14.1 \\
  59468 &  07 27 25.5 & -51 24  09 & 263.15 & -15.64 &  6.71 & 44.48 &G6.5V &  -11 &   22 &  24 &$\pm$ 10 &  58.0 &$\pm$ 11.2 \\
  62644 &  07 42 57.1 & -45 10 23 & 258.58 & -10.58 &  5.04 & 44.50 &G8IV-V&  -20 &  14 &  24 &$\pm$  6 &  72.8 &$\pm$ 7.4 \\
  62848 &  07 43 21.5 & -52  09 51 & 264.96 & -13.74 &  6.68 & 33.71 &F9V   &  -21 &   24 &  32 &$\pm$ 11 &  65.9 &$\pm$  9.3 \\
  63008 &  07 44 12.5 & -50 27 24 & 263.45 & -12.84 &  6.63 & 33.24 &F9V   &  -46 &   37 &  59 &$\pm$ 11 &  70.8 &$\pm$  5.1 \\
  69655 &  08 15 25.2 & -52  03 37 & 267.35 &  -9.41 &  6.62 & 36.96 &G1V   &   16 &  -16 &  23 &$\pm$ 15 & 158.0 &$\pm$ 17.0 \\
  70060 &  08 18 33.3 & -36 39 33 & 254.78 &  -0.42 &  4.40 & 34.93 &A8V   &  -16 &   37 &  41 &$\pm$  7 &  56.5 &$\pm$  4.8 \\
            \noalign{\smallskip}
  71243 & 08 18 31.6 & -76 55 11 & 289.86 & -21.68 &  4.05 & 51.12 &F5V   &  30 &   30 &  43 &$\pm$ 14 &  22.5 &$\pm$  8.7 \\
  73524 &  08 37 20.0 & -40  08 52 & 259.75 &   0.53 &  6.55 & 36.18 &G0Vp  &  -35 &   77 &  85 &$\pm$ 12 &  57.3 &$\pm$  4.2 \\
  77370 &  08 59 24.2 & -59  05  01 & 276.79 &  -8.57 &  5.16 & 38.18 &F4V   &  -36 &    5 &  36 &$\pm$  6 &  86.4 &$\pm$  5.0 \\
  82241 &  09 29 28.6 & -44 31 57 & 269.43 &   4.81 &  6.97 & 24.07 &F8V   &  -31 &   50 &  59 &$\pm$ 11 &  60.9 &$\pm$  5.1 \\
  84117 &  09 42 14.4 & -23 54 56 & 256.70 &  21.52 &  4.94 & 67.47 &F9V   &   14 &   33 &  35 &$\pm$  7 &  33.4 &$\pm$  5.2 \\
  86629 &  09 58 52.3 & -35 53 28 & 267.94 &  15.00 &  5.22 & 30.06 &F1V   &  -18&   15 &  24 &$\pm$  10 &  69.6 &$\pm$ 11.9 \\
  91324 & 10 31 21.8 & -53 42 56 & 283.04 &   3.66 &  4.89 & 45.61 &F9V   &    2 &   13 &  13 &$\pm$  7 &  41.5 &$\pm$ 14.5 \\
  93372 & 10 44 27.0 & -72 26 37 & 293.60 & -11.91 &  6.26 & 31.10 &F6V   &   22 &   -6 &  22 &$\pm$ 12 & 172.2 &$\pm$ 14.5 \\
 101614 & 11 41 26.2 & -41  01 06 & 288.96 &  19.95 &  6.87 & 28.80 &G0V   &  -23 &    5 &  23 &$\pm$ 11 &  84.2 &$\pm$ 12.3 \\
 101805 & 11 42 14.9 & -75 13 38 & 298.47 & -12.96 &  6.47 & 29.35 &F8V   &  216 &   57 & 223 &$\pm$ 11 &   8.7 &$\pm$  1.4 \\
 104731 & 12  03 39.6 & -42 26  03 & 293.60 &  19.57 &  5.15 & 40.44 &F5V   &  -11 &   10 &  15 &$\pm$  9 &  69.8 &$\pm$ 14.8 \\
 \noalign{\smallskip} 
\hline                  
\end{tabular}
\end{table*}

\subsection{Broad-band (400-800 nm) polarization}

Tables 6-8 list the broad-band polarizations computed by weighted averaging of the normalized
Stokes parameters $q$ and $u$ obtained in the $B$, $V$, and $R$ bands at the UH88, T60, and H127 
telescopes, respectively. Statistically significant ($> 3 \sigma$)  polarizations are detected in 115 out of 
the 361 stars observed. 

It is interesting to note the differences in the fraction of  $> 3 \sigma$ detections at the different
telescopes. At the UH88 about one-third of the stars (43 of 125) show measurable polarization, whereas
at the T60 the fraction is about one-quarter (49 of 205). This may be partially due to the fact that
at UH88 we generally observed somewhat fainter stars at larger distances, and the effects from
interstellar dust are therefore probably stronger. When  identifying the target stars, the
fainter magnitudes achievable with UH88 also made it possible to pick more
reddened stars for a given distance compared to stars selected for T60.
The data from H127 show the largest fraction of polarized
stars, namely approximately two-thirds (23 of 31).  This clearly indicates that the average dust content is  larger
in the direction of the H127 sample of stars (Southern hemisphere, 
$ 04^h < \alpha < 12^h, -77\degr < \delta < -4\degr$). 

Figures 3 and 4 show polarization maps based on the broad-band data plotted in the Galactic $(l,b)$ and Equatorial
coordinates. The general view may be slightly complicated, but there are several interesting features 
showing well-aligned polarization vectors. Some of them  resemble "arc" or "loop"-like structures. One  
prominent example can be seen in the aforementioned region of the H127 sample of southern 
stars ($200\degr < l < 315\degr, -75\degr < b < 0\degr$). Further observations would be useful to establish the 
polarization features in this interesting region in greater detail. We also note that the uncertainty of the
instrumental polarization determination is somewhat larger for the H127 sample, particularly for the second 
run (Table 5), due to significant amounts  of polarization in most of the observed stars.
 
When comparing our map with earlier published data on nearby star polarization by \citet{Cotton17},
we can see similarities in the position angle patterns, particularly near the longitudes $l \sim 0\degr$
and $l =260\degr - 315\degr$, even though the samples of observed stars
are entirely different and also extend in the study of these authors to longer distances ($d >$ 50 pc) than in our data.

   \begin{figure}
  \centering
   \includegraphics[width=\hsize]{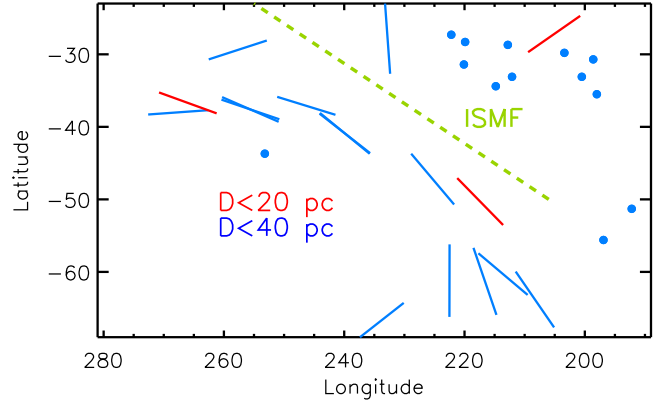}
      \caption{Polarization position angles (bars) of a magnetic filament that is roughly
parallel to the interstellar magnetic field shaping the heliosphere are plotted
for stars within 20 (red) and 40 (blue) parsecs. Stars with $p/e_p$ less than 1.95 are
plotted as dots. The green dashed line indicates the direction of the ISMF
shaping the heliosphere as determined from the center of the IBEX ribbon of
ENAs (see Sect. 3.2). Data are plotted in Galactic coordinates $(l,b)$. 
              }
         \label{Fig5}
   \end{figure}
%

The polarization position angle patterns (Figs. 3 and 4) suggest that the
polarizing dust grains are aligned along magnetic filaments, some with
spatial extents greater than 90$\degr$. Combined with data from several
literature sources, \citet{Frisch2015} found that the position angles for
stars in the longitude range $l = 315\degr - \ 60\degr$ reveal a filament
in the direction toward the heliosphere nose region, as defined by the flow
of interstellar neutral gas through the heliosphere ($l \sim
4\degr, b \sim 15\degr$). Our data (Fig. 3) also clearly show a filament in the
opposite direction ($l \sim 180\degr,b \sim -15\degr$) corresponding to the 
(spatially broader) region of the heliosphere tail.

High-sensitivity measurements of the polarizations of nearby stars provide
the only methodology capable of connecting the ambient ISMF with the
ISMF shaping the heliosphere \citep{Frisch2015}.  The
interstellar magnetic field shaping the heliosphere was discovered to be
traced by a "ribbon" of energetic neutral atoms (ENAs) formed by
charge-exchange between interstellar neutral hydrogen atoms and plasma near the heliopause that 
separates solar and interstellar plasma \citep{McComas09}.  Modeling of the magnetic field creating the
IBEX ribbon \citep{Zirnstein2016} yields an ISMF direction 
toward $l = 26\degr, \ b = 50\degr$. The newly discovered
filament (Figs. 3 and 5) that is centered near $l = 240\degr, \ b= -42\degr$ roughly follows
the direction of the IBEX ribbon ISMF.  Several stars in this filament are within
20 pc, and the nearest star, \object{HD 33262}, is 12 pc away. These new polarization
data firmly place the filament within the cluster of local interstellar clouds
flowing past the Sun and extending to the solar location \citep{Frisch2011}.  
The new filament is also aligned with the projected interface between two of 
these local clouds, Dorados and the Blue Cloud, which have boundaries parallel to each other and  
the polarization filament, and have quasi-perpendicular velocities through the Local Standard of Rest 
(Frisch et al. 2020, in preparation).

Efforts to evaluate the correspondence between polarization strengths and
interstellar column densities for nearby stars have been unsuccessful because
of the low column densities and unknown levels of ionization in the gas
\citep{Frisch15}.  Extended regions with stars lacking significant polarizations
are also found, which can indicate very low dust content, nearby depolarization
screens, or regions at the poles of the magnetic field. 

The star \object{HD 83683} in the T60 data shows much stronger polarization than the others:
 $P(\%)=0.0602 \pm 0.0011,  \theta_{eq}=125.2\degr \pm 0.5\degr$.
This star is in the direction of a known nearby dust cloud ($l \sim 220\degr, b \sim 43\degr$) inside the 
Local Bubble \citep{Meyer06, Peek11}.  Fitting the Serkowski law \citep{Serkowski73} to the   
polarization data in the $B, V$, and $R$
bands yields:  $p_{max} = 0.0649\pm  0.0008$ \%, $\lambda_{max} = 0.706\pm  0.017\mu$m.
Polarization peaking in the red wavelengths indicates size distribution of scattering particles which are 
somewhat greater than the average in the interstellar medium.

The star \object{HD 101805} in the H127 data (Table 8) shows larger polarization than other stars in the sample
($p=0.0223\pm0.0011$\%), most likely due to a contribution of circumstellar origin (Sect. 3.3). 

In the UH88 data, the star \object{HD 126679} ($l \sim 335\degr, b \sim 42\degr$) shows higher 
polarization, $P=0.0595 \pm 0.0008$ \%, $\theta_{eq}=16.4\degr \pm 0.4\degr $, but the wavelength 
dependence looks normal interstellar ($\lambda_{max}= 0.501 \pm 0.060\mu$m, $p_{max}=0.0623 \pm 0.0012 \%$).

Figure 6 shows the dependence of observed polarization with distance. Due to the fact that the degree of polarization is always
positive, for small polarizations the observational errors bias the polarization towards higher values. Statistically,
this can be corrected by $p_c = \sqrt{p^2 - \epsilon^2}$, where $p$ is the observed degree of polarization
and $\epsilon$ its error \citep[see ][]{Serkowski62}. For $p<\epsilon$, we adopt $p_c=0$. The upper boundary of 
the points in Fig. 6 gives a dependence of maximum polarization with distance, $p\ d^{-1} \sim$ 2.9 ppm pc$^{-1}$. 
A linear fit to all points gives  $p\ d^{-1} = 0.37 \pm 0.14 $  ppm pc$^{-1}$. This is much less than
the value $1.64  \pm 0.30$ ppm pc$^{-1}$ derived by \citet{Cotton17} from a sample of southern hemisphere stars.  
The majority of observed points in Fig. 6 lie well below the upper boundary, indicating very low contents
of interstellar dust. There are large regions on the sky with no detectable polarization (see also Figs. 3 and 4) up to distances $d=40-50$ pc. 
The viewing angle between the ISMF and the line of sight also contributes to the low observed
polarization.

   \begin{figure}
   \centering
   \includegraphics[width=\hsize]{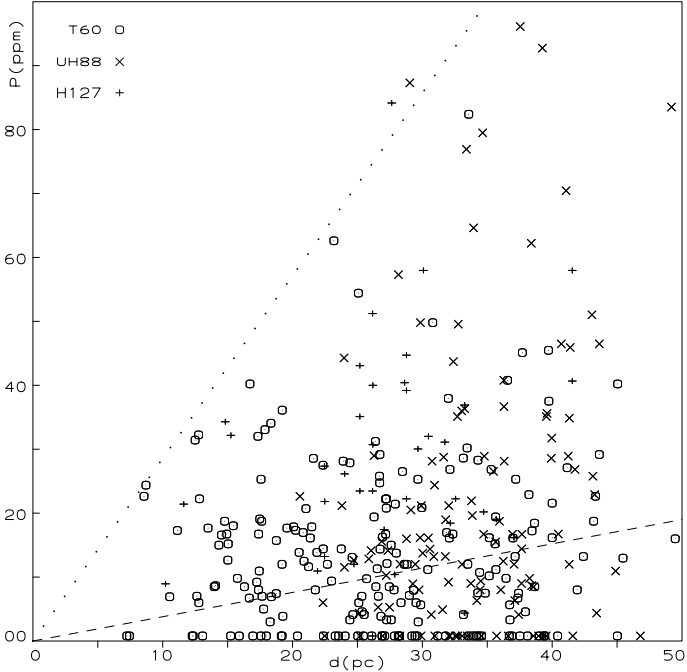}
      \caption{Observed degree of polarization, $p_c$, corrected for the positive bias (see Sect. 3.2), 
     plotted vs. distance, $d$. Upper boundary gives the relation, $p_c \sim 2.9$ ppm pc$^{-1}$, and linear fit 
    to all points (dashed line),  $p_c = 0.37\pm0.14$ ppm pc$^{-1}$.
              }
         \label{Fig5}
   \end{figure}
%

\subsection{Evidence of intrinsic effects from wavelength dependence and/or variability of polarization  }

Three-color ($BVR$) polarimetric data obtained for the star \object{HD 101805} (Table 9)
show polarization increasing from $p = 0.0172 \pm 0.0019$ \%\ in the $R$-band
to $p = 0.0314 \pm 0.0018$ \%\ in the $B$-band. This suggests a contribution from small scattering
particles (Rayleigh-type scattering) in a circumstellar dust and/or gas envelope. Further studies of the circumstellar
environment of HD 101805 would be desirable.

For the majority of the nearby stars ($d<50$ pc) observed, the degree of polarization in the $B, V,$ and $R$
passbands is very low, and constraining the properties of interstellar or circumstellar dust by the wavelength 
dependence is difficult.

%
\begin{table}
\caption{ Evidence of intrinsic polarization in the star HD 101805 from a peculiar wavelength dependence.
The polarization increases from the red towards the blue. The normalized Stokes parameters, $q$ and $u$, 
and the degree of polarization, $p$, are given in units of $10^{-6}$ (ppm), and the position angle 
$\theta$ in the Equatorial frame of reference. }      
\label{table8}      
\centering                          
\begin{tabular}{ccccrc}        
\hline\hline                 
            \noalign{\smallskip}
Filter & $q $  & $u $ & $p \pm e_p$ &$ \theta \pm e_\theta  (\degr)$ & J.D.\\    
\hline                        
  & & & & & \\    
  $B$ &  306 &  72 & 314 $\pm$ 18 &  6.6 $\pm$ 1.7 & 7799.1965 \\  
  $V$ &  175 &  70 & 188 $\pm$ 19 & 10.8 $\pm$ 2.9 & 7799.1965 \\
  $R$ &  164 &  52 & 172 $\pm$ 19 &  8.8 $\pm$ 3.2 & 7799.1965 \\
  & & & & & \\ 
\hline                                   
\end{tabular}
\end{table}
%
%
\begin{table}
\caption{ Stars with multiple observations from the H127 telescope. 
The normalized Stokes parameters, $q$ and $u$, and the degree of polarization, $p$, are given 
in units of $10^{-6}$ (ppm), and the position angle $\theta$ in the Equatorial frame of reference. }      
\label{table9}      
\centering                          
\begin{tabular}{cccllc}        
\hline\hline                 
            \noalign{\smallskip}
HD& $q$ & $u$ & $p \pm  e_p$ & $\theta \pm e_\theta (\degr)$ &  J.D. \\
  \noalign{\smallskip}
\hline
  \noalign{\smallskip}
  29992 &    7 &  -38 &   40 $\pm$   8 &  140.0 $\pm$   5.4 & 7777.9542 \\
  29992 &   10 &  -43 &   45 $\pm$   5 &  141.3 $\pm$   3.1 & 7787.9810 \\
  29992 &    3 &  -22 &   24 $\pm$   9 &  138.3 $\pm$  10.5 & 8134.9991 \\
 &    8 &  -38 &   40 $\pm$   7 &  140.8 $\pm$   5.0 &   av.\\
  \noalign{\smallskip} 
  32743 &   21 &  -11 &   24 $\pm$   5 &  164.9 $\pm$   5.3 & 7778.0907 \\
  32743 &   41 &  -32 &   52 $\pm$   9 &  160.7 $\pm$   4.8 & 7781.9548 \\
 &   26 &  -16 &   31 $\pm$  16 &  163.3 $\pm$  13.5 &  av. \\
  \noalign{\smallskip} 
  50223 &  -29 &   20 &   36 $\pm$   8 &   73.2 $\pm$   6.0 & 7775.0941 \\
  50223 &  -20 &   10 &   24 $\pm$   5 &   77.3 $\pm$   6.3 & 7782.8461 \\
  50223 &  -32 &   30 &   44 $\pm$   9 &   68.9 $\pm$   5.9 & 8125.9939 \\
 &  -24 &   16 &   30 $\pm$   8 &   73.9 $\pm$   7.4 &   av.\\
  \noalign{\smallskip} 
  62644 &  -14 &    6 &   16 $\pm$   9 &   79.5 $\pm$  14.6 & 7778.0816 \\
  62644 &  -21 &   17 &   28 $\pm$   6 &   71.0 $\pm$   6.5 & 7799.0168 \\
 &  -19 &   14 &   24 $\pm$   6 &   72.8 $\pm$   7.4 &   av.\\
  \noalign{\smallskip} 
  77370 &  -40 &    0 &   41 $\pm$   9 &   90.6 $\pm$   6.5 & 7781.1073 \\
  77370 &  -29 &   10 &   32 $\pm$   9 &   81.2 $\pm$   7.6 & 7782.1288 \\
&  -35 &    5 &   36 $\pm$   6 &   86.4 $\pm$   5.0 &  av.  \\
  \noalign{\smallskip} 
  86629 &  -25 &   28 &   38 $\pm$   9 &   66.5 $\pm$   6.3 & 7778.1918 \\
  86629 &   -9 &    3 &   10 $\pm$   9 &   81.2 $\pm$  20.5 & 7782.1653 \\
 &  -17 &   15 &   24 $\pm$  10 &   69.6 $\pm$  11.9 &   av.\\
  \noalign{\smallskip} 
\hline                                   
\end{tabular}
\end{table}

We observed several stars on more than one night. Multiple observations give the possibility to
look for variability on long time intervals from weeks to years. As an initial approach
we have adopted a criterium for picking up candidates for variable polarization: we choose stars 
that show a standard deviation of the normalized Stokes parameters $q$ and $u$ of more than twice  what
is expected from the errors of the nightly points. Table 11 lists measurements of stars that show give
evidence of variable polarization from the  data obtained at the T60 telescope. 
The majority of our multiple observations were made with T60. 

From the total of 205 stars observed at T60 there are 18 stars showing a standard deviation of $q$ and $u$
of more than twice what is expected from the errors of the nightly points (Table 11). A few stars show  
($> 3 \sigma$) night-to-night changes. The clearest examples are \object{HD 19373}, \object{HD 58855}, 
and \object{HD 205289}, where either the  $q$
or $u$ parameter is found to deviate significantly from the  mean values at 
the 2-3$\times 10^{-5}$ level. Still, the evidence for variability is rather marginal, and polarization variations 
exceeding  $10^{-5}$ do not appear to be common in inactive F-G main sequence stars.  More observations
are needed to find evidence for possible periodicity or other systematic time variability.

In our data from the H127 telescope there are five stars observed more than once (Table 10). None of these stars
show statistically significant deviations of $q$ and $u$ from their mean values, that is, variations are $< 2\sigma$. 

We have additional polarization data from observations made at the WHT telescope, and at the NOT telescope, 
at ORM, La Palma. The copy of Dipol-2 polarimeter used at the WHT is the one currently at H127. 
At the NOT a new  version (Dipol-UF) is used. This instrument is equipped with high-speed readout EMCCD 
cameras; otherwise the instrument principle is similar to that of Dipol-2. 

The star \object{HD 6715} shows somewhat stronger polarization, $p(\%)=0.0036 \pm 0.0009$, in our
 measurements at the WHT (Table 12) than observed at the UH88 (Table 6),  $p$(\%)=0.0007 $ \pm$ 0.0009, indicating 
possible intrinsic effects. The polarized star \object{HD 132307} observations at WHT and UH88 are in agreement within the errors. 

The stars observed at the NOT (Table 13) were already measured at the T60 telescope (Table 7).
The results are in good agreement. The star \object{HD 191195} gives a $>3\sigma$ detection at both telescopes: 
$p(\%)=0.0042\pm0.0010, \theta=80.7\degr \pm7.0\degr$, and $p(\%)=0.0039\pm0.0008, \theta=68.2\degr \pm5.6\degr$,
at the T60 and the NOT,  respectively.

    \begin{figure*}
  \centering
 \includegraphics[width=15.5cm]{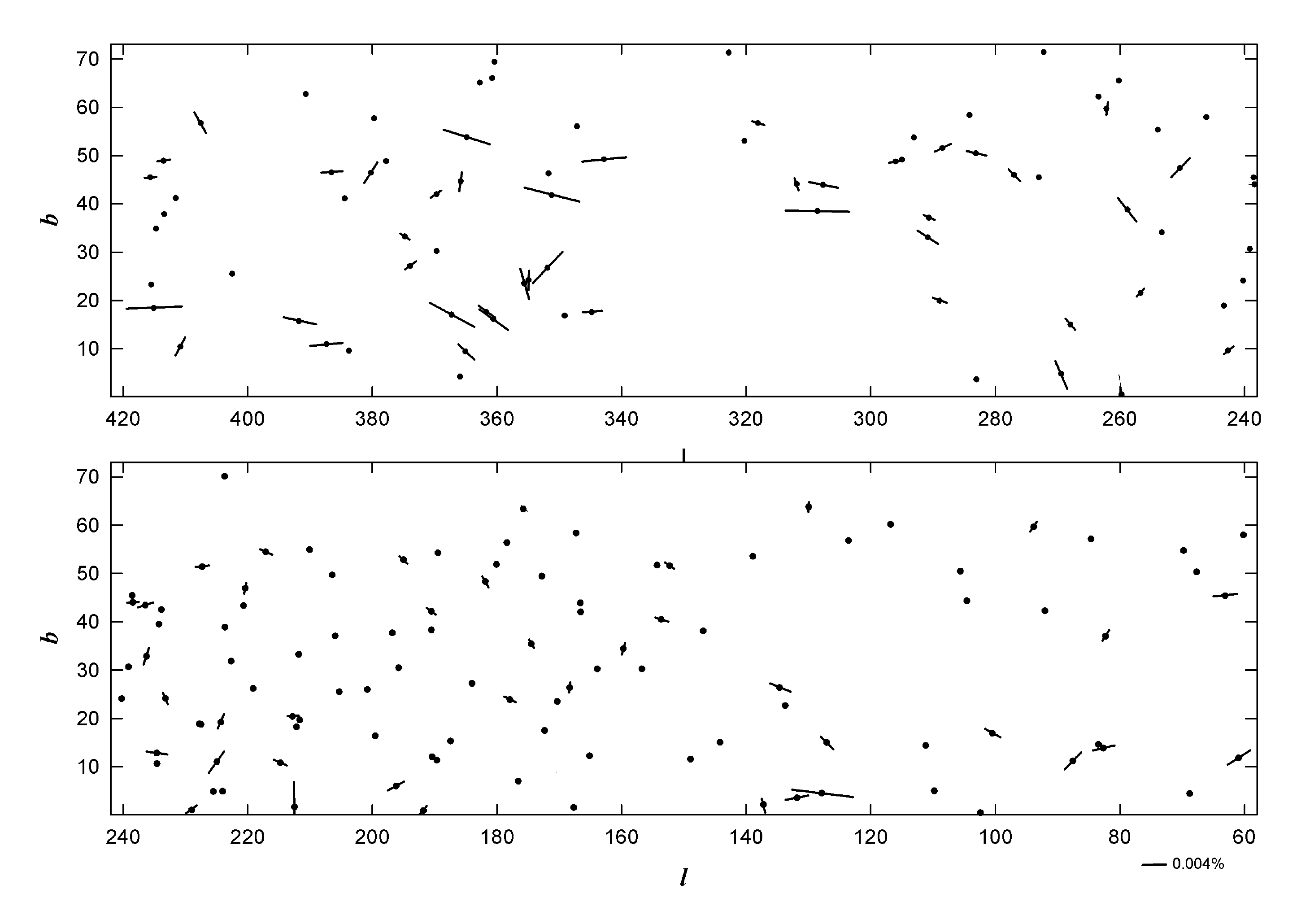}
       \caption{Polarization map of northern Galactic latitude ($b>0\degr$) nearby stars ($d<50$pc), shown for comparison  
with earlier published results of more distant stars (Fig. 8). The polarizations for the peculiar stars HD 83683 and 
HD 126679 are not shown here.
              }
          \label{Fig6}
    \end{figure*}
    \begin{figure*}
  \centering
 \includegraphics[width=15.5cm]{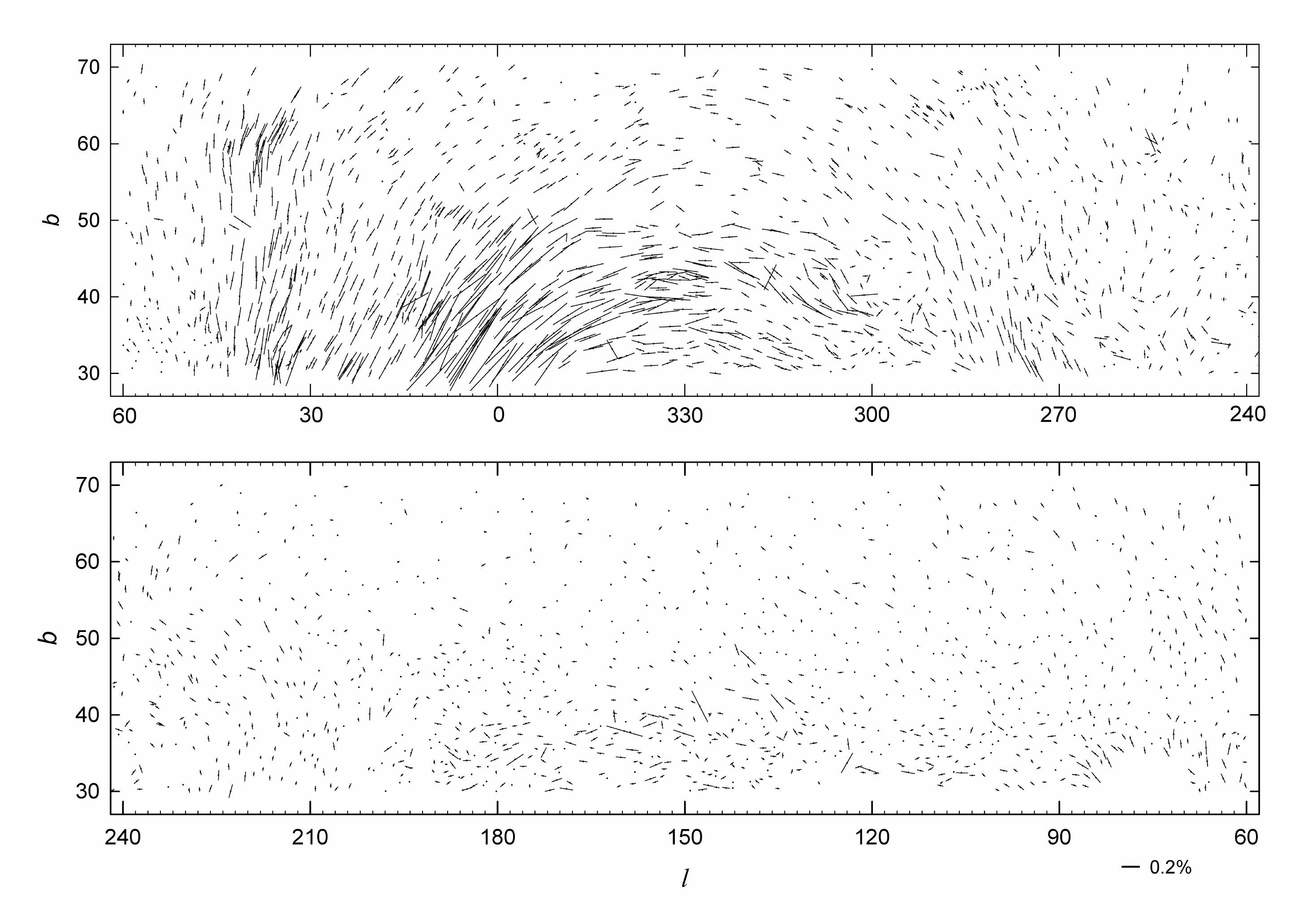}
       \caption{Polarization map of northern Galactic latitude stars in the distance range 100 pc $<d<$ 500 pc. We note that the 
      polarization scale is compressed by a factor of $\sim$70 compared with that in Fig, 7, and
        only stars with $b>30\degr$ are plotted \citep[from][]{Berdyugin14}. 
              }
          \label{Fig7}
    \end{figure*}
 
%

%
\setcounter{table}{11}
\begin{table}
\caption{ Broad-band (400-800 nm) polarimetric data of nearby stars observed at the WHT telescope. 
The normalized Stokes parameters, $q$ and $u$, and the degree of polarization, $p$, are given 
in units of $10^{-6}$ (ppm), and the position angle $\theta$ in the equatorial frame of reference. }      
\label{table9}      
\centering                          
\begin{tabular}{rrrcrc}        
\hline\hline                 
            \noalign{\smallskip}
HD& $q$ & $u$ & $p \pm  e_p$ & $\theta \pm e_\theta (\degr)$ &  J.D. \\
  \noalign{\smallskip}
\hline
  \noalign{\smallskip} 
   6715 &    7 &   35 &   36 $\pm$ 9 &   39.7 $\pm$   6.8 & 7407.3626 \\
  18144 &    8 &   -1 &    8 $ \pm$11 &  178.1 $\pm$27.4 & 7407.4020 \\
  42182 &   -8 &   27 &   28 $\pm$12 &   53.4 $\pm$11.2 & 7407.4430 \\
  51219 &  -19 &  -32 &   38 $\pm$11 &  119.6 $\pm$   8.1 & 7407.4797 \\
  65629 &    7 &  -13 &   14 $\pm$10 &  149.4 $\pm$17.1 & 7407.5174 \\
  77278 &   -7 &   11 &   13 $\pm$   9 &   60.4 $\pm$17.2 & 7407.5700 \\
 117860 &    4 &   16 &   16 $\pm$   5 &   37.9 $\pm$   8.2 & 7206.3865 \\
 132307 &   58 &   18 &   61 $\pm$   8 &    8.5 $\pm$   3.8 & 7207.3826 \\
 140667 &  -27 &    8 &   28 $\pm$   8 &   81.3 $\pm$   8.1 & 7208.3825 \\
 145229 &  -18 &    5 &   19 $\pm$14 &   82.2 $\pm$18.6 & 7209.3942 \\
 150433 &  -21 &    9 &   23 $\pm$   6 &   79.1 $\pm$   6.8 & 7206.4277 \\
 225261 &  -21 &  -38 &   43 $\pm$   9 &  120.6 $\pm$   5.6 & 7407.3097 \\ 
  \noalign{\smallskip} 
\hline                                   
\end{tabular}
\end{table}
\begin{table}
\caption{ Broad-band (400-800 nm) polarimetric data of nearby stars observed at the NOT telescope. 
The normalized Stokes parameters, $q$ and $u$, and the degree of polarization, $p$, are given 
in units of $10^{-6}$ (ppm), and the position angle $\theta$ in the equatorial frame of references. }      
\label{table9}      
\centering                          
\begin{tabular}{rrrrrc}        
\hline\hline                 
            \noalign{\smallskip}
HD& $q$ & $u$ & $p \pm  e_p$ & $\theta \pm e_\theta (\degr)$ &  J.D. \\
  \noalign{\smallskip}
\hline
  \noalign{\smallskip} 
   4813 &    2 &   -6 &    6 $\pm$   7 &  142.2 $\pm$  25.2 & 8687.6996 \\
 132052 &   -3 &   10 &   11 $\pm$   9 &   53.0 $\pm$  19.2 & 8687.3744 \\
 132254 &   10 &   -8 &   13 $\pm$   8 &  160.3 $\pm$  15.5 & 8688.3977 \\
 142373 &   12 &   19 &   22 $\pm$   9 &   28.7 $\pm$  11.0 & 8688.3748 \\
 147449 &    6 &    3 &    7 $\pm$   9 &   15.2 $\pm$  26.2 & 8688.4173 \\
 187013 &  -21 &  -17 &   27 $\pm$   7 &  109.3 $\pm$   7.6 & 8688.5784 \\
 187691 &  -15 &  -12 &   19 $\pm$   6 &  109.6 $\pm$   8.6 & 8687.5400 \\
 191195 &  -28 &   27 &   39 $\pm$   8 &   68.2 $\pm$   5.6 & 8687.5638 \\
 205289 &   -3 &    4 &    5 $\pm$   8 &   65.8 $\pm$  29.1 & 8687.5828 \\
 207958 &   -8 &   -9 &   12 $\pm$   8 &  114.0 $\pm$  16.4 & 8687.6414 \\
 213558 &    3 &   -4 &    5 $\pm$   6 &  156.5 $\pm$  25.4 & 8688.7006 \\
 215648 &  -16 &  -22 &   27 $\pm$   7 &  116.7 $\pm$   7.0 & 8687.6184 \\
 218470 &    1 &   -6 &    6 $\pm$   7 &  140.6 $\pm$  23.0 & 8687.6590 \\
 219623 &   13 &   -9 &   15 $\pm$   7 &  162.4 $\pm$  12.3 & 8688.7211 \\
 225003 &   -5 &  -21 &   21 $\pm$   6 &  128.1 $\pm$   8.1 & 8687.6780 \\ 
  \noalign{\smallskip} 
\hline                                   
\end{tabular}
\end{table} 
%

\subsection{Comparison of polarization maps at high northern Galactic latitudes for the nearby and distant stars}

Figure 7 gives a polarization map for the nearby stars at  northern Galactic latitudes, for comparison with earlier results for more 
distant stars \citep{Berdyugin14}. The most prominent feature seen in the northern ($b > 30^{\circ}$) high-latitude polarization
map for the distant ($d \ge$ 100 pc) stars is a giant "arc" or "loop" between the longitudes $270^{\circ}$ and $45^{\circ}$, 
with the center at $l = 330^{\circ}$ (Fig. 8). 
There is a striking difference between the longitude range 
$l = 240^{\circ} -(360^{\circ}) - 60^{\circ}$ (top panel of Fig. 8), where polarizations are strong and well aligned,
and the range $l = 60^{\circ} - 240^{\circ}$ (bottom panel of Fig. 8), where polarizations are much smaller and no clear 
alignment patterns are visible.  
 
The number of observed nearby stars in Fig. 7 is small, but comparison with Fig. 8 shows similar features: larger polarizations 
in $l = 240^{\circ} -(360^{\circ}) - 0^{\circ}$, and smaller values in $l = 60^{\circ} - 240^{\circ}$.
There is also evidence for the "arc" structure seen in the top panel. Accordingly, local magnetic field structures seen in the 
distance range 100  -- 500 pc  also appear at closer distances. This result is
consistent with a position of the Sun inside the Local Bubble where more nearby
interstellar dust is found in the Galactic center hemisphere than in the
anti-center hemisphere \citep{Frisch17}.

\section{Conclusions}

We collected an extensive high-S/N polarization dataset of 361 nearby stars ($d<50$ pc).
Polarization maps based on these data show patterns of aligned polarization vectors, 
correlating with known nearby dust clouds. 

 The polarization position angles show that the very 
 local ISMF is arranged into distinct magnetic filaments, some with spatial extents greater than 90$\degr$. 
 These magnetic filaments provide a new perspective on the structure of local interstellar clouds that are 
 historically identified primarily by cloud kinematics. There are large regions 
 on the sky with no detectable polarizations ($p<10^{-5}$), up to distances $d=40-50$ pc, which
 indicates very low dust content in these areas, particularly on the northern sky. A linear fit to our sample gives 
 a relation for the average dependence of the degree of polarization versus distance, $p\ d^{-1}=0.37\pm0.14$ ppm pc$^{-1}$.
 This is smaller by a factor of about four than values found for representative regions in the southern
hemisphere \citep{Cotton17}. However, the scatter is large and the values for individual stars in our dataset 
are in the range $0<p\ d^{-1}<2.9$ ppm pc$^{-1}$. Beyond this, there are a few outliers. The extreme case
is  \object{HD 83683}  which is located in a known nearby dust cloud and has $ p\ d^{-1}=14.8\pm0.2$ ppm pc$^{-1}$.
  
 From long-term multiple observations a number 
 ($\sim$ 20) of stars show marginal evidence of intrinsic variability at the $10^{-5}$ level. Three stars show
 statistically significant ($> 3 \sigma$) night-to-night changes. These can
 be attributed to circumstellar effects (e.g., debris disks, chromospheric activity). The star \object{HD 101805}   
 shows a peculiar wavelength dependence with a steep gradient, indicating size distribution of scattering
 particles different from that of typical interstellar medium.
 
Comparison of polarization maps at the high northern Galactic latitudes for distant (100 pc $<d<$ 500 pc) 
and nearby stars ($d < 50$ pc) reveals similar features of polarization patterns, that is, local magnetic field 
structures seen in the distance range $d$ = 100  -- 500 pc also extend to and appear at closer distances. 
Our high-S/N  measurements of intrinsically inactive F-G stars in the magnitude range
$3.8 < V < 9.1$, with very low polarization, also provide a useful dataset for calibration purposes.
 

\begin{acknowledgements}
      This work was supported by the ERC Advanced Grant Hot-
Mol ERC-2011-AdG-291659 (www.hotmol.eu). Dipol-2 was built in the cooperation 
between the University of Turku, Finland, and the Kiepenheuer Institut
f\"{u}r Sonnenphysik, Germany, with the support by the Leibniz Association grant
SAW-2011-KIS-7. We are grateful to the Institute for Astronomy, University of
Hawaii for the observing time allocated for us on the UH88 and T60 telescopes, and
to the University of Tasmania (UTAS) for the observing time at the H127 telescope. 
The Nordic Optical Telescope (NOT), and the William Herschel Telescope (WHT)
are operated in the Spanish Observatorio del Roque de los Muchachos
(ORM) of the Instituto de Astrophysica de Canarias (IAC). Greenhill Observatory is supported by 
the University of Tasmania Foundation. The H127 telescope is funded in part by grant LE110100055 
from the Australian Research Council. The work of P. Frisch was funded by the IBEX mission as 
part of NASA's Explorer Program (80NSSC18K0237) and the IMAP mission as a part of NASA's Solar 
Terrestrial Probes Program (80GSFC19C0027).
\end{acknowledgements}

%
 \bibliographystyle{aa} 
  \bibliography{ref_polar} 

\begin{thebibliography}{23}
\expandafter\ifx\csname natexlab\endcsname\relax\def\natexlab#1{#1}\fi

\bibitem[{{Bailey} {et~al.}(2015){Bailey}, {Kedziora-Chudczer}, {Cotton},
  {Bott}, {Hough}, \& {Lucas}}]{Bailey15}
{Bailey}, J., {Kedziora-Chudczer}, L., {Cotton}, D.~V., {et~al.} 2015, \mnras,
  449, 3064

\bibitem[{{Bailey} {et~al.}(2010){Bailey}, {Lucas}, \& {Hough}}]{Bailey10}
{Bailey}, J., {Lucas}, P.~W., \& {Hough}, J.~H. 2010, \mnras, 405, 2570

\bibitem[{{Berdyugin} {et~al.}(2016){Berdyugin}, {Piirola}, {Sadegi},
  {Tsygankov}, {Sakanoi}, {Kagitani}, {Yoneda}, {Okano}, \&
  {Poutanen}}]{Berdyugin16}
{Berdyugin}, A., {Piirola}, V., {Sadegi}, S., {et~al.} 2016, \aap, 591, A92

\bibitem[{{Berdyugin} {et~al.}(2018){Berdyugin}, {Piirola}, {Sakanoi},
  {Kagitani}, \& {Yoneda}}]{Berdyugin18}
{Berdyugin}, A., {Piirola}, V., {Sakanoi}, T., {Kagitani}, M., \& {Yoneda}, M.
  2018, \aap, 611, A69

\bibitem[{{Berdyugin} {et~al.}(2014){Berdyugin}, {Piirola}, \&
  {Teerikorpi}}]{Berdyugin14}
{Berdyugin}, A., {Piirola}, V., \& {Teerikorpi}, P. 2014, \aap, 561, A24

\bibitem[{{Cotton} {et~al.}(2017){Cotton}, {Marshall}, {Bailey},
  {Kedziora-Chudczer}, {Bott}, {Marsden}, \& {Carter}}]{Cotton17}
{Cotton}, D.~V., {Marshall}, J.~P., {Bailey}, J., {et~al.} 2017, \mnras, 467,
  873

\bibitem[{{Cotton} {et~al.}(2019){Cotton}, {Marshall}, {Frisch},
  {Kedziora-Chudzer}, {Bailey}, {Bott}, {Wright}, {Wyatt}, \&
  {Kennedy}}]{Cotton19}
{Cotton}, D.~V., {Marshall}, J.~P., {Frisch}, P.~C., {et~al.} 2019, \mnras,
  483, 3636

\bibitem[{{Frisch} \& {Dwarkadas}(2017)}]{Frisch17}
{Frisch}, P. \& {Dwarkadas}, V.~V. 2017, {Effect of Supernovae on the Local
  Interstellar Material}, 2253

\bibitem[{{Frisch} {et~al.}(2015{\natexlab{a}}){Frisch}, {Andersson},
  {Berdyugin}, {Piirola}, {Funsten}, {Magalhaes}, {Seriacopi}, {McComas},
  {Schwadron}, {Slavin}, \& {Wiktorowicz}}]{Frisch2015}
{Frisch}, P.~C., {Andersson}, B.~G., {Berdyugin}, A., {et~al.}
  2015{\natexlab{a}}, \apj, 805, 60

\bibitem[{{Frisch} {et~al.}(2015{\natexlab{b}}){Frisch}, {Berdyugin},
  {Piirola}, {Magalhaes}, {Seriacopi}, {Wiktorowicz}, {Andersson}, {Funsten},
  {McComas}, {Schwadron}, {Slavin}, {Hanson}, \& {Fu}}]{Frisch15}
{Frisch}, P.~C., {Berdyugin}, A., {Piirola}, V., {et~al.} 2015{\natexlab{b}},
  \apj, 814, 112

\bibitem[{{Frisch} {et~al.}(2011){Frisch}, {Redfield}, \&
  {Slavin}}]{Frisch2011}
{Frisch}, P.~C., {Redfield}, S., \& {Slavin}, J.~D. 2011, \araa, 49, 237

\bibitem[{{Hsu} \& {Breger}(1982)}]{Hsu82}
{Hsu}, J.~C. \& {Breger}, M. 1982, \apj, 262, 732

\bibitem[{{McComas} {et~al.}(2009){McComas}, {Allegrini}, \&
  {Bochsler}}]{McComas09}
{McComas}, D.~J., {Allegrini}, F., \& {Bochsler}, P. e.~a. 2009, Science, 326,
  959

\bibitem[{{Meyer} {et~al.}(2006){Meyer}, {Lauroesch}, {Heiles}, {Peek}, \&
  {Engelhorn}}]{Meyer06}
{Meyer}, D.~M., {Lauroesch}, J.~T., {Heiles}, C., {Peek}, J.~E.~G., \&
  {Engelhorn}, K. 2006, \apjl, 650, L67

\bibitem[{{Peek} {et~al.}(2011){Peek}, {Heiles}, {Peek}, {Meyer}, \&
  {Lauroesch}}]{Peek11}
{Peek}, J.~E.~G., {Heiles}, C., {Peek}, K. M.~G., {Meyer}, D.~M., \&
  {Lauroesch}, J.~T. 2011, \apj, 735, 129

\bibitem[{{Piirola}(1973)}]{Piirola73}
{Piirola}, V. 1973, \aap, 27, 383

\bibitem[{{Piirola} {et~al.}(2014){Piirola}, {Berdyugin}, \&
  {Berdyugina}}]{Piirola14}
{Piirola}, V., {Berdyugin}, A., \& {Berdyugina}, S. 2014, in \procspie, Vol.
  9147, Ground-based and Airborne Instrumentation for Astronomy V, 91478I

\bibitem[{{Schmidt} {et~al.}(1992){Schmidt}, {Elston}, \& {Lupie}}]{Schmidt92}
{Schmidt}, G.~D., {Elston}, R., \& {Lupie}, O.~L. 1992, \aj, 104, 1563

\bibitem[{{Schwadron} {et~al.}(2009){Schwadron}, {Bzowski}, \&
  {Crew}}]{Schwadron09}
{Schwadron}, N.~A., {Bzowski}, M., \& {Crew}, G.~B. 2009, Science, 326, 966

\bibitem[{{Serkowski}(1962)}]{Serkowski62}
{Serkowski}, K. 1962, Advances in Astronomy and Astrophysics, 1, 289

\bibitem[{{Serkowski}(1973)}]{Serkowski73}
{Serkowski}, K. 1973, in IAU Symposium, Vol.~52, Interstellar Dust and Related
  Topics, ed. J.~M. {Greenberg} \& H.~C. {van de Hulst}, 145

\bibitem[{{Turnshek} {et~al.}(1990){Turnshek}, {Bohlin}, {Williamson}, {Lupie},
  {Koornneef}, \& {Morgan}}]{Turnshek90}
{Turnshek}, D.~A., {Bohlin}, R.~C., {Williamson}, R.~L., I., {et~al.} 1990,
  \aj, 99, 1243

\bibitem[{{Zirnstein} {et~al.}(2016){Zirnstein}, {Heerikhuisen}, {Funsten},
  {Livadiotis}, {McComas}, \& {Pogorelov}}]{Zirnstein2016}
{Zirnstein}, E.~J., {Heerikhuisen}, J., {Funsten}, H.~O., {et~al.} 2016, \apjl,
  818, L18

\end{thebibliography}
%

 
 

\clearpage
\onecolumn

\setcounter{table}{5}

\begin{longtable}{rcrrrrrlrrrlrl}
\caption{\label{UH88data} Broad-band (400-800 nm) polarimetric data of nearby stars observed
   at the UH88 telescope. The normalized Stokes parameters, $q$ and $u$, and the degree of polarization, 
  $p$, are given in units of $10^{-6}$ (ppm), and the position angle $\theta$ in the equatorial frame of reference. } \\
\hline\hline
  \noalign{\smallskip}
HD &  R.A. & Decl.    & $l(\degr)$  &  $b(\degr)$ & $V_{mag}$  &  Par. & Sp & $q$ &  $u$ & $p$ &$\pm e_p$ & $\theta(\degr)$ &$\pm e_\theta (\degr)$  \\    
  \noalign{\smallskip}
\hline
\endfirsthead
\caption{continued.}\\
\hline\hline
  \noalign{\smallskip}
HD &  R.A. & Dec & $l(\degr)$  &  $b(\degr)$ & $V_{mag}$  &  Par. & Sp & $q$ &  $u$ & $p$ &$\pm e_p$ & $\theta(\degr)$ &$\pm e_\theta (\degr)$  \\ 
  \noalign{\smallskip}
\hline
\endhead
            \noalign{\smallskip}
    101 &  00 05 54.7 &  18 14  06 & 108.00 & -43.31 &  7.46 & 26.72 &F8V   &   -6 &   -7 &   9 &$\pm$  8 & 114.7 &$\pm$ 20.6 \\
    361 &  00 08 16.4 & -14 49 28 &  81.53 & -74.12 &  7.03 & 36.57 &G1V   &  -15 &   -1 &  15 &$\pm$  5 &  92.7 &$\pm$  2.4 \\
   1562 &  00 20 00.4 &  38 13 39 & 116.17 & -24.23 &  6.99 & 40.46 &G1V   &   14 &   -4 &  14 &$\pm$  6 & 172.5 &$\pm$  7.2 \\
   1832 &  00 23 00.2 &  22 22 30 & 114.34 & -40.02 &  7.59 & 24.57 &F8V   &  -15 &  -45 &  47 &$\pm$  7 & 125.7 &$\pm$  4.4 \\
   4208 &  00 44 26.7 & -26 30 57 &  53.97 & -88.32 &  7.79 & 29.22 &G7V   & -105 & -115 & 156 &$\pm$  9 & 113.8 &$\pm$  1.7 \\
   4915 &  00 51 10.8 &  -05 02 21 & 122.76 & -67.91 &  6.97 & 44.48 &G5V   &    1 &   10 &  10 &$\pm$ 11 &  42.8 &$\pm$ 23.1 \\
   5035 &  00 52 40.2 &  31 27 34 & 123.24 & -31.41 &  8.36 & 24.71 &G0V   &  -16 &  -10 &  19 &$\pm$  9 & 106.5 &$\pm$ 12.4 \\
   5294 &  00 54 59.3 &  24 06 01 & 123.97 & -38.76 &  7.38 & 33.43 &G5V   &   19 &  -14 &  23 &$\pm$  9 & 161.8 &$\pm$ 10.2 \\
\end{longtable}
 
%
%

\begin{longtable}{rcrrrrrlrrrlrl}
\caption{\label{T60data} Broad-band (400-800 nm) polarimetric data of nearby stars observed
   at the T60 telescope. The normalized Stokes parameters, $q$ and $u$, and the degree of polarization, 
  $p$, are given in units of $10^{-6}$ (ppm), and the position angle $\theta$ in the Equatorial frame of references. } \\
\hline\hline
  \noalign{\smallskip}
HD &  R.A. & Decl.    & $l(\degr)$  &  $b(\degr)$ & $V_{mag}$  &  Par. & Sp & $q$ &  $u$ & $p$ &$\pm e_p$ & $\theta(\degr)$ &$\pm e_\theta (\degr)$  \\    
  \noalign{\smallskip}
\hline
\endfirsthead
\caption{continued.}\\
\hline\hline
  \noalign{\smallskip}
HD &  R.A. & Dec & $l(\degr)$  &  $b(\degr)$ & $V_{mag}$  &  Par. & Sp & $q$ &  $u$ & $p$ &$\pm e_p$ & $\theta(\degr)$ &$\pm e_\theta (\degr)$  \\ 
  \noalign{\smallskip}
\hline
\endhead
%
            \noalign{\smallskip}
    123 &  00  06 15.8 &  58 26 12 & 117.03 &  -3.92 &  5.98 & 46.56&  G5V   &   19 &    6 &  20 &$\pm$  9 &   8.3 &$\pm$ 12.6 \\
    203 &  00  06 50.1 & -23  06 27 &  52.21 & -79.14 &  5.74 & 25.39&  F3V   &   -4 &    1 &   4 &$\pm$  9 &  82.1 &$\pm$ 32.0 \\
    693 &  00 11 15.9 & -15 28  05 &  82.23 & -75.07 &  4.89 & 53.34&  F8V   &  -12 &  -10 &  16 &$\pm$  3 & 109.1 &$\pm$  4.9 \\
   1388 &  00 17 58.9 & -13 27 20 &  91.52 & -74.25 &  6.50 & 36.74&  G0V   &  -19 &   14 &  24 &$\pm$ 12 &  71.8 &$\pm$ 13.6 \\
   3795 &  00 40 32.8 & -23 48 18 &  85.85 & -85.87 &  6.14 & 35.13&  K0V   &  -15 &   23 &  28 &$\pm$  9 &  61.6 &$\pm$  9.2 \\
   4247 &  00 44 44.4 & -22  00 22 & 106.02 & -84.66 &  5.22 & 37.39&  F3V   &  -16 &    1 &  16 &$\pm$  7 &  87.5 &$\pm$ 11.8 \\
   4307 &  00 45 28.7 & -12 52 51 & 117.05 & -75.68 &  6.15 & 31.13&  G0V   &    7 &   27 &  28 &$\pm$  8 &  37.9 &$\pm$  8.0 \\
   4628 &  00 48 23.0 &   05 16 50 & 121.51 & -57.58 &  5.74 &134.50&  K2.5V &   -2 &    3 &   4 &$\pm$  8 &  59.2 &$\pm$ 32.8 \\
\end{longtable}

\setcounter{table}{10}
\begin{longtable}{ccccrc}
\caption{\label{kstars} Stars showing evidence of polarization variability from the observations at the T60 telescope. 
  The normalized Stokes parameters,
   $q$ and $u$, and the degree of polarization, $p$, are given in units of $10^{-6}$ (ppm) for stars showing
observed standard deviation of $q$ and $u$ $>$ twice the value expected from the errors of the nightly points.}\\
\hline\hline
  \noalign{\smallskip}
HD& $q$ & $u$ & $p \pm  e_p$ & $\theta \pm e_\theta (\degr)$ &  J.D. \\
  \noalign{\smallskip}
\hline
\endfirsthead
\caption{continued.}\\
\hline\hline
  \noalign{\smallskip}
HD& $q$ & $u$ & $p \pm  e_p$ & $\theta \pm e_\theta (\degr)$ &  J.D.   \\
  \noalign{\smallskip}
\hline
\endhead
%
  \noalign{\smallskip}
  16673 &  -21 &   31 &   37 $\pm$   9 &   61.7 $\pm$   7.0 & 8092.8290 \\
  16673 &   -8 &   -6 &   10 $\pm$   8 &  107.7 $\pm$  19.1 & 8428.9127 \\
 av. &  -14 &   10 &   17 $\pm$  15 &   71.5 $\pm$  21.0 &  \\
  \noalign{\smallskip} 
  19373 &   -9 &    5 &   10 $\pm$   4 &   75.2 $\pm$  11.3 & 7041.7720 \\
  19373 &    0 &    7 &    7 $\pm$   7 &   43.0 $\pm$  22.2 & 7299.0752 \\
  19373 &   -3 &   12 &   12 $\pm$   9 &   51.6 $\pm$  19.0 & 7777.7679 \\
  19373 &  -10 &   44 &   45 $\pm$   8 &   51.5 $\pm$   4.9 & 8027.1158 \\
 av. &   -7 &   12 &   13 $\pm$  11 &   60.2 $\pm$  19.7 &  \\
\end{longtable}

\end{document}